# A Program for Integrating Math and Physics Internet-Based Teaching Tools into Large University Physics Courses

*David Toback*[1,§], *Andreas Mershin*[1,2,#], *Irina Novikova*[1,3,♦]

*[1] Department of Physics, Texas A&M University, College Station, TX 77840*

*[2] Center for Biomedical Engineering, Massachusetts Institute of Technology, Cambridge, MA 02139, USA*

*[3] Harvard-Smithsonian Center for Astrophysics, Cambridge, MA 02138*

**Abstract**

Significant obstacles prevent large, university-level, introductory physics courses from effectively teaching problem-solving skills. We describe our program for integrating three internet-based "teaching-while-quizzing" tools to address two of these barriers: students' poor math skills and instructors' insufficient grading recourses. We outline our system of math remediation, homework and after-homework quizzes, and mini-practice exams, and demonstrate how it can be incorporated into courses with modest instructor effort.

[§] toback@physics.tamu.edu

[#] mershin@mit.edu

[♦] inovikova@cfa.harvard.edu

**Introduction**

Teaching students in our large, introductory, calculus-based physics courses [1] to be good problem-solvers is a difficult task. Not only must students be taught to understand and use the physics concepts in a problem, they must become adept at turning the physical quantities into symbolic variables, translating the problem into equations, and "turning the crank" on the mathematics to find both a closed-form solution and a numerical answer. Physics education research has shown that students' poor math skills and instructors' lack of pen-and-paper homework grading resources, two problems we face at our institution, can have a significant impact on problem-solving skill development [2, 3, 4]. While Interactive Engagement methods appear to be the preferred mode of instruction [5], for practical reasons we have not been able to widely implement them. In this article, we describe three internet-based "teaching-while-quizzing" tools we have developed and how they have been integrated into our traditional lecture course in powerful, but easy to incorporate, ways [6]. These are designed to remediate students' math deficiencies, automate homework-grading, and guide study time towards problem-solving. Our intent is for instructors who face similar obstacles to adopt these tools that are available upon request [7].

Web-based instruction systems and their advantages have been discussed for some time [2, 8]. Although some research indicates that simply collecting and grading homework online provides no measurable advantage over high-quality, fast-turnaround human grading [9], other studies show that using computer tools for repeated quizzing and remediation can have long-term advantages over voluntary, no-feedback homework assignments [4, 10]. To create our program we relied on both our teaching experience and the considerable literature on good physics teaching practice and psychology, especially the studies that emphasize explicit goals, immediate and constant feedback, balance between skill-level and challenge, time-on-task, high expectations, motivation for performance, and opportunity to repeat until mastery [11, 12, 13]. While other programs have incorporated these ideas (see for example [10, 14]), there are subtle but significant differences in our implementation and integration that allow instructors to incorporate our methods into the course with modest effort.

**Overview of the tools and how they are incorporated**

Our tools consist of three separate sets of short quizzes administered via the internet and thus require no substantive extra work for instructors to incorporate. They are: i) the Automated Mathematics Evaluation System (AMES); ii) the Computerized Homework Assignment Grading System (CHAGS); and iii) a set of after-homework QUizzes Intended to Consolidate Knowledge (QUICK).

The topics within the quizzes and the requirements to pass them are a central part of what makes them powerful. Following Mastery Learning [12] (repetition until achieving a certain score) and Precision Teaching [13] (repetition until achieving a predetermined number of correct answers per unit time) motifs, we require students to score a perfect 100% on each quiz, within the allotted time, in order to move on to the next quiz. If they fail, we indicate the correct answers and give them an unlimited number of attempts (without penalty) to get a perfect score, changing the problems on each attempt. This way, we combat students' temptation to ignore the harder problems. Crucially, quizzes are tailored such that multiple topics and difficulty levels are covered each time, encouraging them to learn the material as a whole.

The sequencing of the quizzes is also important. The course begins with the AMES math quizzes, and students are not allowed to move on until they have mastered all of them. After finishing AMES, a student moves on to CHAGS for chapter 1 homework, and then QUICK for chapter 1 homework quizzes. Again, a student must get a perfect score on all questions in order to move on to CHAGS chapter 2, and so on. Achieving a perfect score on all the relevant materials before each in-class, pen-and-paper exam opens an extra-credit, QUICK mini-practice exam. Students must complete AMES, CHAGS and QUICK to pass the course. We next discuss each tool and how their incorporation achieves its aim.

**Automated Mathematics Evaluation System (AMES)**

The AMES quizzes are strictly limited to the relevant pre/co-requisite math topics for the course and are assigned during the first week of the semester. Our intent is to remind students of the relevant calculation tools they should already possess and to have them practice until they have (re)gained facility. We encourage them to drop the course if they cannot perform well quickly, since there is otherwise a good chance they will fail [3]. Each quiz consists of ten multiple-choice problems, developed by the authors, which target the most common math deficiency areas of our students. Every attempt at an AMES quiz includes at least one problem from the following areas: i) simple algebraic expressions in one variable; ii) systems of equations in two variables; iii) quadratic equations and identities; iv) geometry and trigonometry including vectors; v) fractions, numbers, exponents, powers of ten; vi) word problems and proportionalities; and vii) simple differentiation and integration. Students have ten minutes to complete each quiz, and they must obtain a 100% on ten "separate" quizzes (although each quiz is actually drawn randomly from the same pool of questions). Completion of AMES effects any necessary remediation early in the semester.

**Computerized Homework Assignment Grading System (CHAGS)**

For homework, students receive a list of problems from the primary textbook and are expected to solve each in closed form, with pen and paper, *before* accessing CHAGS. There, the same problems are presented on screen, but with different numeric parameters, and students are required to quickly type in the new numeric answer. They have only a short time (20 minutes) to substitute the new number(s) into their formulae, so they are forced to be ready with their closed form solutions before attempting submission. Numerical answers encourage students to be careful in their calculations, and they fosters both unit and "reality" checks for the realistically possible magnitudes of various physical quantities. The time restriction emphasizes the utility of closed-form solutions and the need to do the work before logging on to the computer.

**Quizzes Intended to Consolidate Knowledge (QUICK)**

To reinforce learning after students have achieved a 100% on their homework, we have implemented a ten-minute, two-problem, multiple-choice quiz for each textbook chapter using problems taken from a standard textbook test bank [15]. Each QUICK consists of intermediate-level problems randomly drawn from a large pool, with new problems drawn for each attempt. After successfully completing the QUICKs for the chapters covered on the upcoming exam, students earn access to a voluntary four-problem, twenty-minute mini-practice exam (again repeatable without penalty) which provides immediate feedback on their level of preparedness; extra credit is given for a 100% before the exam. This encourages time-on-task problem-solving preparation for the exam, shown in the literature to be significant [16], and reinforces keeping up with the course.

**Discussion and Conclusions**

The majority of our tools have been widely adopted by most instructors at our institution while fewer than 3 years ago, there was no web-based learning in our department.

Our preliminary evidence and informal anonymous student questionnaires indicate that the systems are helpful. For example, the bulk of the students readily work within the systems, are more likely to keep up with the course content (typically >60% get the voluntary mini-practice exam extra credit), spend considerably more time on problem-solving, and do not waste time on less helpful practices such as equation memorization or multiple scans of the textbook. We have also noticed a behavioral shift in our weaker students towards symbol-based problem-solving methods and self-checking. Instructors using the system have reported a marked decrease in student complaints about difficulty following simple mathematical steps and have noticed an increased general readiness for exams. While we do not yet have a formal, quantitative assessment of our methods, the literature indicates that our combination of remediation, Mastery/Precision and quizzing methods is likely to be effective [4, 10, 16].

In summary, we have developed an easy-to-implement program of administering internet-based quizzes to develop math skills, check homework and offer practice-problem feedback for introductory physics courses. Our methods have been pervasively adopted at our own institution, and we hope future controlled studies will more precisely assess their apparent benefits.

**Acknowledgments**

We would like to thank Wayne Saslow, Teruki Kamon, Peter McIntyre, Bob Webb, George Welch, Cathy Ezrailson, Nancy Simpson, David Meltzer, and Joan Wolf for useful discussions, and Joel Walker, Matt Cervantes, Rhonda Blackburn, Jim Snell, Court Samson, Jonathan Asaadi and Sally Yang for help with programming and other technical expertise. AM was partially supported by the S.A. Onassis Public Benefit Foundation.

**References**

[1] These are "traditional" lecture courses on classical mechanics and electricity & magnetism taught at Texas A&M University with typically 1,500 students per semester (mostly first year engineers) between the two courses. Students are broken into lectures of ~120 and further subdivided into recitations of ~30.

[2] See for example, L. Hsu, E. Brewe, T. M. Foster and K. A. Harper, "Resource Letter RPS-1: Research in Problem Solving," Am. J. Phys. **72**, 1147 (Jan. 2004).

[3] H.T. Hudson and W.R. McIntire, "Correlation Between Mathematical Skills and Success in Physics," Am. J. Phys. **45**, 470 (May 1977). For more recent data see D. Meltzer, "The relationship between mathematics preparation and conceptual learning gains in physics: A possible 'hidden variable' in diagnostic pretest scores," Am. J. Phys. **70** 12 (Dec. 2002).

[4] R. Dufresne, J. Mestre, D.M. Hart and K.A. Rath, "The Effect of Web-Based Homework on Test Performance in Large Enrollment Introductory Physics Courses," Jl. of Comp. in Math and Science Teaching **21**, 229 (Sept. 2002).

[5] See, for example, R. R. Hake, "Interactive-engagement versus traditional methods: A six-thousand-student survey of mechanics test data for introductory physics courses," Am. J. Phys. **66**, 64 (Jan. 1998) and references therein.

[6] This has been a 3-year program with financial support from the Texas A&M University Department of Physics, Instructional Technology Services, and an award from the Montague Scholarship Program at the Center for Teaching Excellence.

[7] While we have implemented these tools using WebCT, they can be readily deployed via many other software packages or in-house written code. For more information about WebCT see http://www.webct.com. To obtain copies of our WebCT zip files for the quizzes and problem database, contact the authors. For a nice list of other delivery systems see Ref. [11].

[8] S. W. Bonham, A. Titus, R. J. Beichner and L. Martin, "Education Research using Web-Based Assessment Systems," J. Research on Computing in Education **33**, 28 (2000).

[9] See S. W. Bonham, D.L. Deardorff, and R. J. Beichner, "Comparison of Student Performance Using Web and Paper-Based Homework in College-Level Physics," *J. Res. Sci. Teach.* **40**, 1050 (Nov. 2003), and S. Bonham, R. Beichner, and D. Deardorff, "On-line homework: does it make a difference?" *Phys. Teach.* **39** 293 (May 2001).

[10] M. J. Marr, E. W. Thomas, M. R. Benne, A. Thomas and R. M. Hume, "Development of Instructional Systems for Teaching and Electricity and Magnetism Course for Engineers," *Am. J. Phys.* **67**, 9 (Sept. 1999).

[11] See, for example T. A. Angelo, "A Teacher's Dozen: Fourteen General, Research-Based Principles for Improving Higher Learning in Our Classrooms," *AAHE Bulletin* **45,** 3 (Apr. 1993).

[12] For a review of Mastery Learning, see D. Davis and J. Sorrell, "Mastery Learning in Public Schools," paper prepared for PSY 702: Conditions of Learning (1995). http://chiron.valdosta.edu/whuitt/files/mastlear.html.

[13] See O. R. Lindsey, "From Skinner to precision teaching," in *Let's Try Doing Something Else Kind of Thing*, edited by J. B. Jordan and L. S. Robbins (Council on Exceptional Children, Arlington VA, 1972), pp 1-12.

[14] J. Risley, "Motivating Students to Learn Physics Using an Online Homework System," *APS Forum on Education*, fall 2001 Newsletter.

[15] We have taken the test item file by E. Oberhofer, D. Curott and R. Pelcovits from "Physics for Scientists and Engineers" by D.C. Giancoli, Prentice Hall, 3rd Edition (2000).

[16]  G. Tang and A. Titus, "Increasing Students' Time on Task in Calculus and General Physics Courses through WebAssign," *Proceedings of the 2002 ASEE Conference and Exposition.*